\documentclass[aip,reprint]{revtex4-1}

\usepackage{graphicx,subfigure,float,amssymb,amsmath,color}
\usepackage{mathdots}
\usepackage{bm}
\begin{document}


\newcommand{\Eqref}[1]{Eq.~(\ref{#1})}
\newcommand{\Eqsref}[2]{Eqs.~(\ref{#1}) and (\ref{#2})}
\newcommand{\refcite}[1]{Ref.~\cite{#1}}
\newcommand{\figref}[1]{Fig.~(\ref{#1})}
\newcommand{\blue}[1]{\textcolor{blue}{#1}}
\newcommand{\red}[1]{\textcolor{red}{#1}}

\title{Josephson effects in an alternating current biased transition edge sensor}



\author{L. Gottardi} 
\email[]{Electronic mail: l.gottardi@sron.nl}
\affiliation{SRON Netherlands Institute for Space Research, Sorbonnelaan 2, 3584 CA Utrecht, The Netherlands}        

\author{A. Kozorezov}
\affiliation{Department of Physics, Lancaster University, LA1 4YB, Lancaster, UK}

\author{H. Akamatsu}
\author{J. van der Kuur}
\author{M.P. Bruijn}
\author{R.H. den Hartog}
\author{R. Hijmering}
\author{P. Khosropanah}
\affiliation{SRON Netherlands Institute for Space Research, Sorbonnelaan 2, 3584 CA Utrecht, The Netherlands}   
\author{C. Lambert}
\affiliation{Department of Physics, Lancaster University, LA1 4YB, Lancaster, UK}                                                  
\author{A.J. van der Linden}
\author{M.L. Ridder}
\author{T. Suzuki}
\affiliation{SRON Netherlands Institute for Space Research, Sorbonnelaan 2, 3584 CA Utrecht, The Netherlands}                                                     
\author{J.R. Gao}
\affiliation{Kavli Institute of NanoScience, Faculty of Applied Sciences, Delft University of Technology, Lorentzweg 1, 2628 CJ Delft, The Netherlands}
\affiliation{SRON Netherlands Institute for Space Research, Sorbonnelaan 2, 3584 CA Utrecht, The Netherlands}



\date{\today}

\begin{abstract}
 We report the  experimental evidence of the ac Josephson effect in a transition edge sensor (TES) operating in a frequency domain multiplexer and  biased by ac voltage at MHz frequencies. The effect is observed by measuring the non-linear impedance of the sensor. The TES is treated as a weakly-linked superconducting system and  within the resistively shunted junction model framework. We provide a full theoretical explanation of the results by finding the analytic solution of the non-inertial  Langevian equation of the system and calculating the non-linear response of the detector to a large ac bias current in the presence of noise. 
\end{abstract}


\maketitle 



%
%

%

Superconducting transition-edge sensors (TESs) are highly sensitive
thermometers widely used as radiation detectors over an energy range from
near infrared to gamma rays. In particular we are developing TES-based detectors for the infrared SAFARI/SPICA\cite{Spica} and the X-ray XIFU/Athena\cite{Athena} instruments. 
TESs are in most cases low impedance devices that operate in
the voltage  bias regime while the current is generally
read-out by a SQUID current amplifier. Both a constant or an
alternating bias voltage
can be used \cite{IrwinHilton,JvdK02}. In  the latter case  changes of the TES resistance
induced by the thermal signal 
modulate the amplitude of the ac bias current. The small signal detector response is
modelled in  great details both under dc and ac bias
\cite{Swetz12,JvdKuur11}.  Those models however do not fully explain
all the physical phenomena recently observed in TESs.
It has been recently demonstrated that TES-based devices behave
as weak-links due to longitudinally induced superconductivity from
the leads via the proximity effect \cite{Sadleir10} and a detailed experimental
investigation of the weak-link  effects in dc biased x-ray
microcalorimeters has been reported \cite{Smith13}. 
Evidence of weak-link effects in ac biased TES microcalorimeters has been given \cite{Gottardi12xray}, but an adequate experimental and theoretical investigation is still missing.
We previously  proposed a theoretical framework \cite{Kozorezov11} based on the resistively shunted junction model (RSJ) that can be used to describe the resistive state of a TES under dc bias. 
 In this letter, we extend the model to calculate the stationary non-linear response of a TES to a large ac bias current in the presence of noise and  we compare it to the experimental data  obtained with a TES-based bolometer. We report  a clear signature of the ac-Josephson effect in the TES biased at MHz frequencies. 

\begin{figure}[ht]
\includegraphics[width=7.25cm]{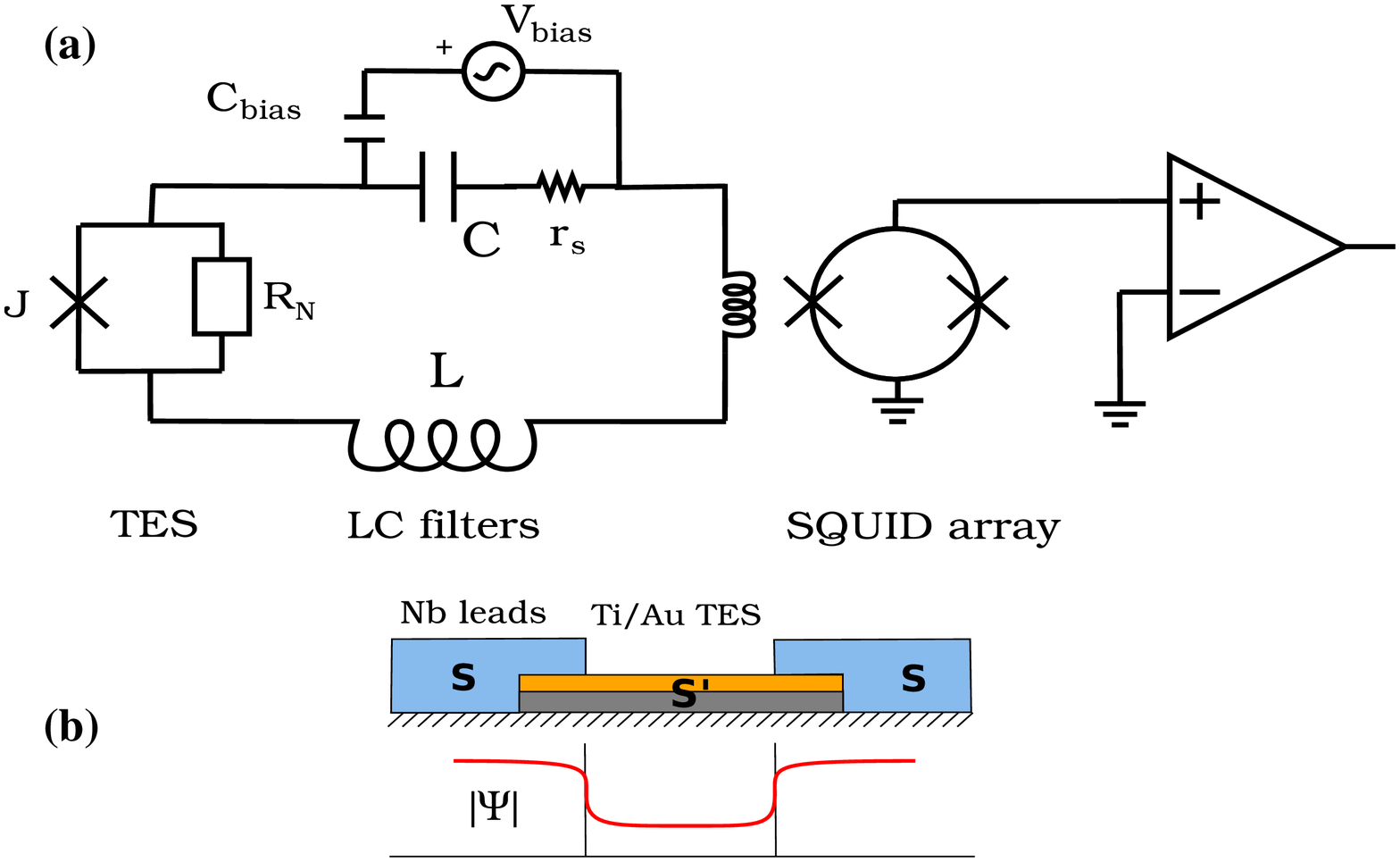}
\caption{\label{fig:scheme} {\bf (a)} Electrical circuit of the ac-bias read-out where the TES is modelled as a resistively shunted junction. {\bf (b)} Cross section of the TES bilayer with Nb leads and the dependence on the superconducting order parameter over the TES length. }
\end{figure}

The general equation for the Frequency Domain Multiplexing (FDM) electrical circuit, simplified for a single resonator is\cite{JvdKuur11}
\begin{equation}
V(t)=I(t)Z_{TES}(T,I(t))+L\frac{dI(t)}{dt}+\frac{1}{C}\int I(t)dt+r_sI(t)
\end{equation}     
\noindent where $V(t)$ is the total voltage across the TES, $L$ and $C$ are respectively the inductance and the capacitance of the bias circuit, $r_s$ is the total stray resistance in the circuit and $Z_{TES}$ is the TES impedance, which depends on temperature $T$ and current $I(t)$.  
As previously reported \cite{Sadleir11,Smith13}, the superconducting leads proximitize the
TES  bilayer film over a distance defined by the coherence length
$\xi$. As a result, the superconducting order parameter $|\Psi|$ is
spatially dependent over the length of the bilayer, as shown in the
cross section of \figref{fig:scheme}(b) . The TES can be seen then  as an
$SS'S$ proximity induced weak-link where the electrical bias leads and the Ti/Au bilayer are the $S$ and $S'$ materials respectively. 

The electrical scheme of the ac bias read-out and the TES representation as a weak-link are shown in \figref{fig:scheme}. We describe the TES using the RSJ model in which the resistance is replaced
by an ideal junction shunted by the TES normal resistance $R_{N}$ \cite{Likharev79,Kozorezov11}. 

Within this model, the total current $I(t)$ flowing between two
weakly connected superconductors can be described by the two Josephson
equations as 
\begin{equation}
\label{acdcJosephson}
I(t)+\mathcal{L}(t)=I_c \sin\varphi+\frac{V(t)}{R_{N}}, \;\;\; \frac{\partial \varphi}{\partial t}=\frac{2e}{\hbar}V(t)
\end{equation}



\noindent where $\varphi$ is the gauge-invariant phase difference
between the wave functions of the two superconductors and  $I_c$ is the
critical current. $\mathcal{L}(t)$ is a thermal fluctuation noise current superimposed on the bias current generated by the fact that the weak-link operates at temperature $T$ above absolute zero \cite{Ambega69,Coffey00}. $\mathcal{L}(t)$ is assumed  to be Gaussian white noise with $\overline{\mathcal{L}(t)}=0$ and $\overline{\mathcal{L}(t)\mathcal{L}(t')}=\frac{2kT}{R_N}\delta(t-t')$.

It follows from the second of the Josephson Eqs.~(\ref{acdcJosephson})
that in a TES under ac bias voltage $V(t)=V_{ac}\cos\omega_0 t$ the
superconducting phase $\varphi$ oscillates at the same bias frequency
$\omega_o$, but $\pi/2$-out-of-phase with respect to the voltage.
The total current flowing in the TES is then  a superposition of a
supercurrent and a normal current
caused by the flow of quasiparticle across the junction.

Within this context we evaluate the stationary  nonlinear response of a TES for a large ac bias current and in the presence of noise.   
For a given ac bias current $I(t)=I_{ac}\cos\omega_0 t$, from Eqs.~(\ref{acdcJosephson}), we can derive $\varphi$ by solving the non-inertial Langevin equation for the RSJ model with noise:
\begin{equation}
\label{eqvarphi}
\frac{\tau}{\gamma}\frac{\partial \varphi}{\partial t}+\sin \varphi=\zeta \cos\omega_0 t+I_c^{-1}\mathcal{L}(t),
\end{equation}

\noindent  where $\tau=\left(\hbar/2e\right)^2/R_{N}k_BT$ is the averaged time a particle takes to diffuse one period of a tilted washing board potential, $\zeta=I_{ac}/I_c$ and $\gamma=\hbar I_c/\left(2ek_BT\right)$ is the normalised Josephson coupling energy, a parameter which also characterises  the noise strength with $\gamma=\infty$ corresponding to the noiseless limit.
The total phase difference across the weak-link  $\varphi$ depends as well on the external magnetic flux coupled into the TES and its expression can be generalised as $\varphi_{tot}=\varphi+2\pi\frac{\Phi(t)}{\Phi_0}$. In the first approximation $\Phi(t)=A_{eff}(B_{\perp,DC}+B_{\perp,AC}(t))$ where $A_{eff}$ is the effective weak-link area and $B_{\perp,DC}$ and $B_{\perp,AC}(t)$ the dc and ac  perpendicular magnetic field crossing the TES respectively.

The solution of \Eqref{eqvarphi} for the stationary state can be  found analytically in terms of matrix continued fraction \cite{Coffey00}. The non-linear admittance of the weakly superconductive TES at the carrier frequency $f_0=\omega_0/2\pi$ can be derived in the form 
\begin{equation}
\label{Zwl}
\frac{1}{Z_{TES}}=\frac{1}{R_{TES}}-j\frac{1}{\omega_0L_J}=\frac{1}{R_N\left( 1-j\zeta\left[c^1_1(\omega)-c^{1*}_1(-\omega)\right]\right)},
\end{equation} 
\noindent with the non-linear Josephson inductance $L_J$ in parallel with the TES resistance $R_{TES}$.

The elements of the vector ${\bf c_n}$ can be calculated from the following equations 
\begin{equation}
\label{cn}
\begin{array}{lll}
{\bf c}_n & =& {\bf S}_n^{21}{\bf c}_0,\\\\
{\bf S}^{21}_n & =& ({\bf q}_{2n-1}+({\bf q}_{2n}+{\bf S}^{21}_{n+1})^{-1})^{-1},
\end{array}
\end{equation}
\noindent where the tridiagonal infinite matrices ${\bf q_n}$  can be written as
\begin{equation}
\label{qnmatrix}
\begin{array}{lll}
 
{\bf q}_n & =&j\left(\begin{matrix}
                    \ddots&\vdots&\vdots&\vdots&\vdots&\vdots&\iddots\\
                    \hdots&z_n^{-2}& \zeta&0&0&0&\hdots\\
                    \hdots&\zeta&z_n^{-1}&\zeta&0&0&\hdots\\
                    \hdots&0&\zeta&z_n^{0}&\zeta&0&\hdots\\
                    \hdots&0&0&\zeta&z_n^{1}&\zeta&\hdots\\
                    \hdots&0&0&0&\zeta&z_n^{2}&\hdots\\
                    \iddots&\vdots&\vdots&\vdots&\vdots&\vdots&\ddots
                    \end{matrix}\right)
\end{array}
\end{equation}
\noindent and  the vector ${\bf c}_0=\left(\hdots\; 0\; 1\; 0\; \hdots \right)^T$. The diagonal elements in \Eqref{qnmatrix} are defined as
\begin{equation}
\label{zkn}
z_n^m(\omega)=2\left(m\omega\tau /\left(n\gamma\right)-j n/\gamma\right),
\end{equation}

\noindent with $m=\{\hdots,-1,0,1,\hdots\}$ and $n=\{1,2,\hdots\}.$




We assume the TES to be  ac biased with the bias  current derived from the temperature and current
dependent resistance $R_{TES}(T,I)$  and the effective power balance equation. 

The FDM read-out of TESs measures naturally the resistive $R_{TES}$
and reactive $X=\omega_oL_{J}$ components of the TES non-linear impedance in
\Eqref{Zwl} by performing the in-phase and quadrature detection of the TES
current. By measuring both the amplitude $I_{ac}$ and the phase $\theta$  of the TES
current with respect to the applied voltage $V_{ac}$ we obtain the TES non-linear impedance $Z_{TES}=V_{ac}/I_{ac}e^{-j\theta}$. The effective power of a TES with resistance $R_{TES}(T,I)$ under ac bias is given by\cite{JvdKuur11}
\begin{equation}
\label{Powereqsteady}
\frac{V_{ac}I_{ac}}{2}\cos \theta=\frac{1}{2}I_{ac}^2 R_{TES}(T,I)\cos^2\theta=k(T^n-T^n_b),
\end{equation}
\noindent where $k=G/n(T^{n-1})$ with $G$ the differential thermal conductance, $n$ is the thermal conductance
exponent and  $T_{b}$ is the bath temperature\cite{IrwinHilton}.

The vectors ${\bf c}_n$ can be calculated numerically by computing the matrix continued fraction in \Eqref{cn} and solving simultaneously the power balance equation \Eqref{Powereqsteady}. The convergence is rapidly achieved for the parameters discussed below.

The set-up used for the experiments presented in this paper is an FDM
system working in the frequency range from 1 to 5 MHz, based on an
open-loop or baseband feedback read-out of a linearised two-stage
SQUID amplifier and high$-Q$ lithographic $LC$ resonators \cite{Gottardi14}.
Here below we report the results for a TES ac biased at frequencies of
$1.4\, \text{and}\, 2.4\, \mathrm{MHz}$.
The device under test is a bolometer based on a square Ti/Au (16/60 nm)
bilayer TES $L\times L = 50\times 50 \, \mu \mathrm{m^2}$ in size.  It has a critical
temperature at zero magnetic field of $T_c=85.5\,\mathrm{mK}$, a  normal state  resistance  of
$R_{N,tes}=98\,\mathrm{m\Omega}$, a thermal conductance to the bath $G=0.3\, \mathrm{pW/K}$ and a
measured NEP under ac bias of $3.5\times 10^{-19}\, \mathrm{W/\sqrt{Hz}}$ \cite{Gottardi14}.
The electrical contact to the bolometer is realised by $90\,\mathrm{nm}$ thick
Nb leads deposited on the edges of the bilayer as shown in
\figref{fig:scheme}. 
We report strong evidence of proximity induced  weak-link behaviour of
our TES-based bolometer measured under ac bias. The three main
experimental 
observations are: the modulation of
the TES critical current $I_c$ under applied magnetic field $B$, showing the
typical Fraunhofer-like pattern of a Josephson junction; the
exponential dependence on the critical current on temperature $T$ 
and the presence of a non-linear reactance modulated by the TES bias
 voltage. 

\begin{figure}
\includegraphics[width=7.cm]{{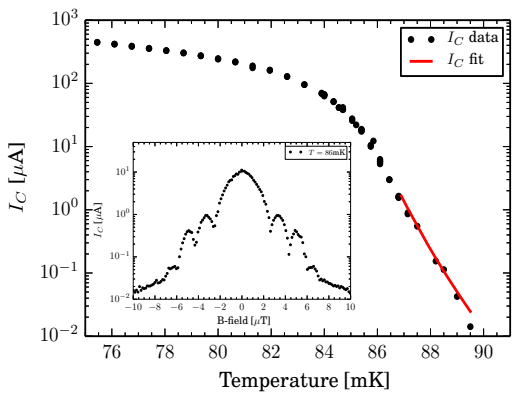}}
\caption{\label{fig:icvstb} TES critical current as a function of the bath temperature at zero magnetic field  and (inset) as a function of the applied perpendicular magnetic field at $T=86\, \mathrm{mK}$. The TES is ac biased at a frequency of $2.4\, \mathrm{MHz}$. The model \cite{Sadleir10,Likharev79} fit the data in the TES $R(T,I)$ sharp transition range (red line).}
\end{figure}

The measured critical current $I_c(T,B)$ is shown in \figref{fig:icvstb} as
a function of temperature $T$ and the applied perpendicular dc  magnetic field $B_{\perp,DC}$. 
For $T\geq T_c$ the critical current of the TES shows an exponential
dependence on $T$ and can be  fitted (see red line in \figref{fig:icvstb}) by the approximated formula $I_{c}(L,T) \propto e^{-\frac{L_{wl}}{\xi_0}|\frac{T}{T_{Ci}}-1|^{1/2}}$ to estimate the Josephson contribution to $I_c$ from the weak-link model \cite{Sadleir10,Likharev79}. Here $L_{wl}\leq L$ is the effective size of the TES along the current flow operating as a
weak-link, $\xi_0$ is the zero temperature coherence length and $T_{Ci}$ is the intrinsic critical  temperature of the laterally unproximised TiAu bilayer.   




The periodicity of the oscillations in the Fraunhofer pattern of 
$I_c(B)$ is defined as $\Delta B_{min}=\Phi_0/LL_{wl}$ where $\Delta B_{min}$ is the magnetic field
difference between the local minima in $I_c(B)$. From the data we
inferred that only a fraction $f=\frac{\Delta B_{min}L^2}{\Phi_0}=\frac{L_{wl}}{L}=0.67$
of the geometrical TES area $L^2$ behaves as a weak-link. The explanation for this experimental result requires a detailed study of the electrodynamics of long junctions and goes beyond the scope of this paper. The fact that the magnetic field penetrates only part of the TES  has implication for the choice of the value of the normal
resistance used in the RSJ model presented below.  
 

The TES resistive state is calculated  by simulating the resistive transition within the RSJ model for dc-biased TES \cite{Kozorezov11} and using as input parameters the experimental critical current curve $I_c(T)$ and the TES normal resistance. The use of the functional dependence $R(T,I)$ calculated for the dc bias case is justified by the our previous results \cite{Gottardi14} where we have shown that the IV curves measured under ac and dc bias  are identical within the experimental errors. 
Because only $67\%$ of the TES area is proximized we
assume the normal resistance of the weak-link to
be  $R_{N}=0.67R_{N,tes}$. 



\begin{figure}[h]
\center
\includegraphics[width=7.25cm]{{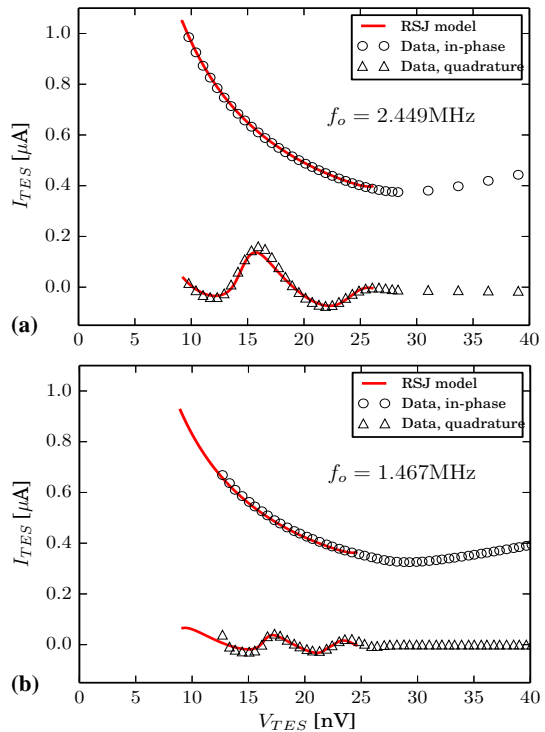}}
\caption{\label{fig:iv2fbias} TES $I-V$ characteristics showing both the in-phase (open circle) and quadrature (open triangle) components of the current. The measurements where taken at a bias frequency of 2.4 ({\bf a}) and $1.4\,\mathrm{MHz}$ ({\bf b}) and at a bath temperature of $30$ and $40\, \mathrm{mK}$ respectively. The solid lines are the $IV$ curves obtained from the RSJ model.}
\end{figure}

The use of the general  formalism of continued fraction matrices to solve \Eqref{eqvarphi} requires self-consistent numerical solution of the balance equation \Eqref{Powereqsteady}. We developed the code which allows to implement this procedure and calculate the in-phase
and quadrature components of the TES current and the non-linear
impedance of  the weakly superconducting TES as  given in \Eqref{Zwl}.
The experimental  in-phase and  quadrature TES current  components are
plotted in \figref{fig:iv2fbias} versus the TES bias voltage for $1.4$
and $2.4\,  \mathrm{MHz}$ bias frequencies. The  curves calculated using
the RSJ model are also shown.
The IV curve obtained using the in-phase
component of the  TES current is consistent with  the results obtained
with   the  same   pixel   measured  under   dc-bias \cite{Gottardi14}.  The quadrature component  of the  current shows
an oscillatory    behaviour   dependent    on    the driving    bias
frequency. Generally, the period and the amplitude  of the oscillations decreases with the
bias frequency. Moreover, the amplitude is larger at low bias voltages due  to the fact that  the noise parameter  $\gamma$ increases when
the TES is biased low in  the transition. Both observations  are  consistent
with  the numerical calculation  performed by  Coffey {\it et al.}\cite{Coffey00} on Josephson junctions. For
the detector under test the $\pi/2$-out-phase current reaches a maximum peak 
amplitude of about $10\%$ and $20\%$ of the total current flowing in the TES for a driving frequency of $1.4\, \mathrm{MHz}$ and $2.4\, \mathrm{MHz}$ respectively.

\begin{figure}[h]
\center
\includegraphics[width=7.25cm]{{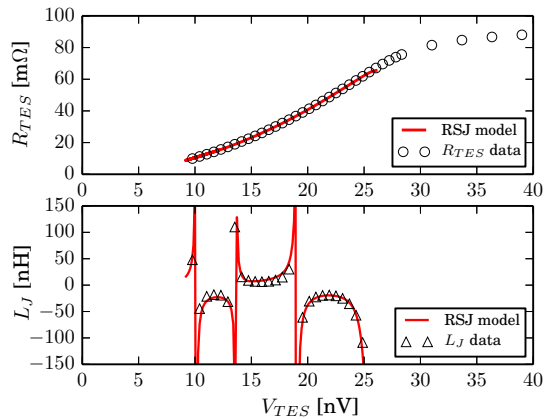}}
\caption{\label{fig:Zwl} The TES resistance $R_{TES}$ (open circle) and the Josephson inductance $L_J$ (open triangle) as a function of the voltage bias. The TES is ac biased at a frequency of 2.4 MHz. The solid lines show the results from the RSJ model.}
\end{figure}

\begin{figure}[h]
\center
\includegraphics[width=7.25cm]{{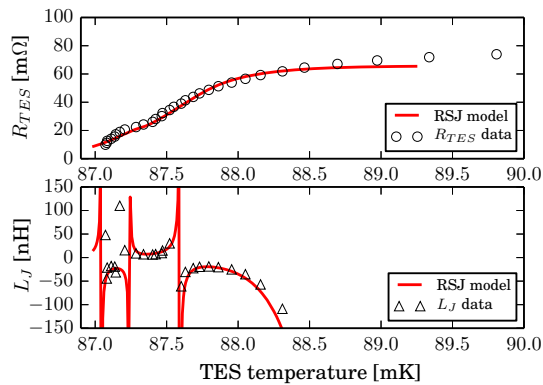}}
\caption{\label{fig:RXT} TES resistance $R_{TES}$ (open circle) and Josephson inductance $L_J$ (open triangle)  as a function of temperature $T$ derived from the $I-V$ curves and the power balance equation. The TES is ac biased at a frequency of 2.4 MHz. The solid lines show the result from the RSJ model.}
\end{figure}

The measured and calculated TES resistance $R_{TES}$
and Josephson inductance $L_{J}$ as a function of the bias voltage and temperature  are shown  respectively in   \figref{fig:Zwl} and \figref{fig:RXT} for  the detector that was  biased at  $f_o=2.4\,\mathrm{MHz}$.
The TES behaves as a nonlinear inductor in parallel with a resistance as predicted by the RSJ model. The nonlinear inductance oscillates between positive and negative values in the superconducting transition. At some particular bias voltage the inductance becomes infinite and the TES is purely resistive.

The $R(T)$ curves of the TES
for $T>T_c$ show typically a
sharp resistive transition above which the resistance is not constant,
but has a small nonzero slope. The RSJ model explains the
shape of the sharp transition, while, as shown by Sadleir {\it et
  al.}\cite{Sadleir11} the enhanced conductivity above the sharp phase
transition can be well fitted under the assumption that zero-resistance region penetrates a distance twice the temperature dependent coherence length of the infinite bi-layer from both leads. 

In conclusion, we observed a clear signature of the ac Josephson effect in a TES bolometer operating under ac bias at frequencies  of few MHz. The effect clearly appears in the quadrature component of the bias current. 
We applied the RSJ model and calculated the stationary non-linear response of a TES under ac bias and  in the presence of noise. 
Using the  analytic expressions for the non-linear admittance of a weakly superconducting TES changing in accordance with the power balance variation through the resistive transition we can fully reproduce the measured TES resistance and the Josephson inductance as a function of bias voltage, bias frequency and operating temperature. 

\begin{acknowledgments}
H.A. is supported by a Grant-in-Aid for Japan Society 
for the Promotion of Science (JSPS) Fellows (22-606).
\end{acknowledgments}

\bibliography{gottardi_weaklink_arx_reviewed}

\end{document}